\documentclass[conference]{IEEEtran}
\IEEEoverridecommandlockouts
\usepackage{cite}
\usepackage{amsmath,amssymb,amsfonts}
\usepackage{algorithmic}
\usepackage{graphicx}
\usepackage{textcomp}
\usepackage{xcolor}
\usepackage{hyperref}
\usepackage{mathtools}
\mathtoolsset{showonlyrefs}

\graphicspath{{img/}{IC_data/}}

\DeclareMathOperator*{\argmin}{arg\,min}
\newcommand{\domain}{\mathcal{M}}
\newcommand{\range}{\mathbb{R}}
\newcommand{\criticalIndex}{\mathcal{I}}
\newcommand{\persistence}{\mathcal{P}}
\newcommand{\persistenceDiagram}[1]{\mathcal{D}(#1)}

\newcommand{\pointMetric}[1]{d}
\newcommand{\liftedMetric}[1]{\widehat{\pointMetric{#1}}}
\newcommand{\wasserstein}[1]{W}
\newcommand{\diagramSpace}{\mathbb{D}}
\newcommand{\isovalue}{w}
\newcommand{\sublevelset}[1]{#1^{-1}_{-\infty}}

\newcommand{\diagramBarycenter}{\mathcal{D}^*}
\newcommand{\diagram}{\mathcal{D}}

\newcommand{\needsVerification}[1]{\textcolor{red}{#1}}
\renewcommand{\needsVerification}[1]{\textcolor{black}{#1}}

\newcommand{\criticalPointMixing}{\lambda}
\newcommand{\geometricLifting}{\alpha}

\newcommand{\removable}[1]{\textcolor{purple}{#1}}
\renewcommand{\removable}[1]{}

\newcommand{\jules}[1]{\textcolor{black}{#1}}

\def\BibTeX{{\rm B\kern-.05em{\sc i\kern-.025em b}\kern-.08em
    T\kern-.1667em\lower.7ex\hbox{E}\kern-.125emX}}
\begin{document}

\title{Statistical Parameter Selection\\for Clustering Persistence Diagrams
 \thanks{This work is partially supported by the European Commission grant H2020-
 FETHPC-2017 ``VESTEC'' (ref. 800904).
 }
}

\author{\IEEEauthorblockN{Max Kontak}
\IEEEauthorblockA{\textit{Simulation and Software Technology} \\
\textit{DLR German Aerospace Center}\\
K\"oln, Germany \\
max.kontak@dlr.de}
\and
\IEEEauthorblockN{Jules Vidal}
\IEEEauthorblockA{\textit{CNRS LIP6} \\
\textit{Sorbonne Universite}\\
Paris, France\\
jules.vidal@sorbonne-universite.fr}
\and
\IEEEauthorblockN{Julien Tierny}
\IEEEauthorblockA{\textit{CNRS LIP6} \\
\textit{Sorbonne Universite}\\
Paris, France\\
julien.tierny@sorbonne-universite.fr}}

\maketitle

\begin{abstract}
In urgent decision making applications, ensemble simulations are an important way to determine different outcome scenarios based on currently available data.
In this paper, we will analyze the output of ensemble simulations by considering so-called persistence diagrams, which are reduced representations of the original data, motivated by the extraction of topological features.
Based on a recently published progressive algorithm for the clustering of persistence diagrams, we determine the optimal number of clusters, and therefore the number of significantly different outcome scenarios, by the minimization of established statistical score functions.
Furthermore, we present a proof-of-concept prototype implementation of the statistical selection of the number of clusters and provide the results of an experimental study, where this implementation has been applied to real-world ensemble data sets.
\end{abstract}

\begin{IEEEkeywords}
urgent decision making, 
ensemble simulation,
topological clustering,
statistical model selection
\end{IEEEkeywords}

\section{Introduction}

To support urgent decision making in the situation of a catastrophic event, ensemble simulations can be used to quantify uncertainties and to distinguish different possible outcome scenarios, which may require diverse steps to be taken by a crisis manager.
\jules{
In practice, modern numerical simulations are subject to a variety of
input parameters, related to the initial conditions of the system under
study, as well as the configuration of its environment. Given recent
advances in hardware computational power, engineers and scientists
can now densely sample the space of these input parameters, in order to
identify the most plausible crisis evolution.
The European project VESTEC \cite{vestec} focuses on building a toolchain 
combining interactive supercomputing, data analysis and visualization for the 
purpose of urgent decision making.
Through the VESTEC system, a crisis manager would be able to run an ensemble 
of numerical simulations and interactively explore the resulting data in order
to help the decision making process. Three use cases are to be supported: 
mosquito-borne diseases, wildfire monitoring, and space weather forecasting.
}

The identification of \jules{the possible} scenarios can be accomplished by 
finding
clusters in the simulation results.
\jules{For instance, for a time-varying wildfire simulation, the outputs of all 
ensemble simulations for each time step could be clustered to obtain a time 
series of clusterings, which can then be further analyzed.
In that way, one can determine points in time at which significantly different 
simulations arise in the ensemble (\emph{e.\,g.}, there is only one fire vs.\ the fire 
has split up into multiple parts), which is relevant for the decision maker, who 
can compare the different clusters with the behavior of the fire in reality to 
identify the most plausible crisis evolution.}

Unfortunately, the output data sets of large-scale simulation codes are often too big to all fit in memory, which creates a need for reduced data representations.
These can be provided by \emph{topological data analysis} \cite{edelsbrunner09, tierny_book}.
So-called \emph{persistence diagrams} have been used in many applications before (combustion 
\cite{laney_vis06, bremer_tvcg11, gyulassy_ev14}, 
fluid dynamics \cite{kasten_tvcg11, fangChen13},
material sciences \cite{gyulassy07, favelier16, Lukasczyk17},
chemistry \cite{chemistry_vis14, harshChemistry}, 
and astrophysics \cite{sousbie11, shivashankar2016felix}) to obtain reduced data representations.
Recently, an efficient technique has been proposed for clustering persistence diagrams instead of the original simulation data \cite{vidal_vis19}.
With the application of urgent decision making in mind, the algorithm has been designed based on the classical $k$-means clustering algorithm to incorporate time constraints.
However, the number of clusters $k$ is still a parameter of the approach in \cite{vidal_vis19}.

In this work, we will investigate so-called \emph{information criteria}, which have been developed for statistical model selection \cite{Claeskens2008,Konishi2007}, to determine the optimal number of clusters.
We will present a proof-of-concept prototype implementation for the statistical selection of parameters in topological clustering and perform an experimental study on real-life ensemble data sets.

\begin{figure*}
    \centering
    \includegraphics[width=\linewidth]{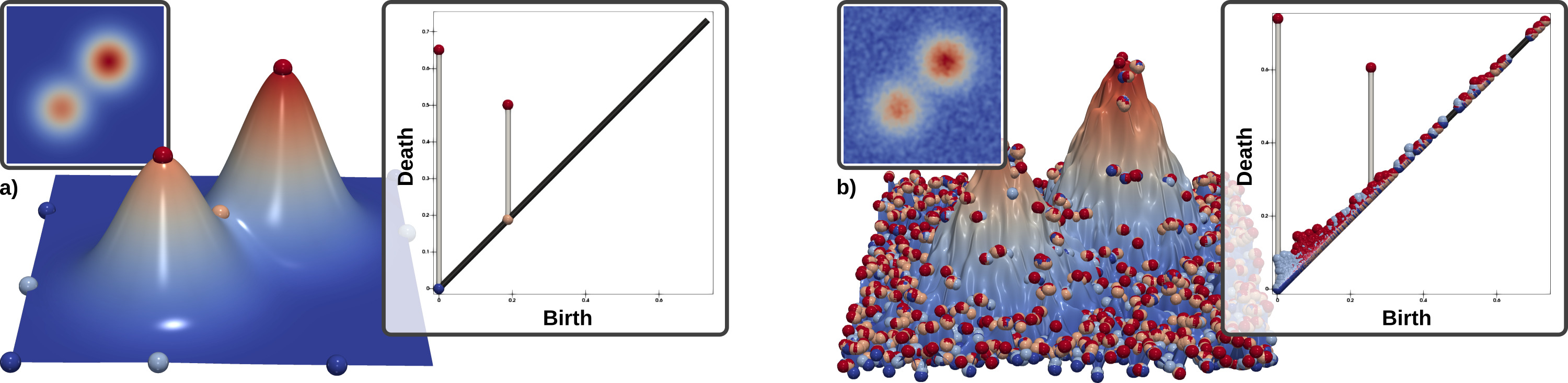}
    \caption{The persistence diagram (right insets) reduces a data set (top 
left, 
        bottom left: terrain view) to a 2D point cloud where each off-diagonal 
point 
        represents a \emph{topological feature}. 
        \jules{In this diagram, the X and Y axes denote the \emph{birth} and 
\emph{death} of a topological feature, respectively.}
        In these examples, points which 
stand 
        out from the diagonal represent large features (the two hills, (a) and 
(b)), 
        while points near the diagonal correspond to noisy features in the 
data.}
    \label{fig:toyExample}
\end{figure*}

\section{Related work}
Existing techniques for ensemble visualization and analysis typically 
construct, for each member of the ensemble, some geometrical objects, such as 
level sets or streamlines, which are used to cluster the original members of 
the ensemble and to construct aggregated views.
Several methods have been proposed,
such as spaghetti plots \cite{spaghettiPlot} for level-set
variability in weather data ensembles \cite{Potter2009,Sanyal2010}, or box-plots
for the variability of contours \cite{whitaker2013} and curves in general
\cite{Mirzargar2014}. 

Related to our work, clustering
techniques have been used to analyze the main trends in ensembles of
streamlines \cite{Ferstl2016} and isocontours \cite{Ferstl2016b}. 
Favelier et al. \cite{favelier2018} introduced an approach, relying on spectral 
clustering, to analyze critical point variability in ensembles.
Lacombe et al. \cite{lacombe2018} introduced an approach to cluster 
ensemble members based on their persistence diagrams. More relevant to our 
context of urgent decision making, Vidal et al. \cite{vidal_vis19} introduced 
a method for the progressive clustering of persistence diagrams, 
supporting   computation time constraints. However, this approach, which 
extends the seminal \emph{k-means} algorithm \cite{lloyd82}, is subject to an 
input parameter, 
the number of output clusters $k$, which is often difficult to tune in practice.

\section{Background}
This section presents the technical background of our work.

\subsection{Topological Data Analysis}
\label{sec_background_tda}
Topological Data Analysis is a recent set of techniques \cite{edelsbrunner09, 
tierny_book}, which focus  on  structural data representations. We review in 
the following the main ingredients for the computation of topological 
signatures of data, for their comparison, and for their clustering. This section 
contains definitions taken from Vidal et al. \cite{vidal_vis19}, 
reproduced here for self-completeness.

\paragraph{Persistence diagrams}
The input data
is an ensemble of $n$ piecewise linear (PL) scalar fields $f : \domain 
\rightarrow \range$ defined on  a PL $d$-manifold $\domain$, with
$d\leq3$ in our applications. 
We note $\sublevelset{f}(\isovalue)=\{p \in \domain~|
~f(p) < \isovalue\}$ the \textit{sub-level set} of $f$. When continuously increasing 
$\isovalue$, the topology of 
$\sublevelset{f}(\isovalue)$ 
can only 
change at specific locations, called the \emph{critical points} of $f$. 
Critical points are classified according to their \textit{index} 
$\criticalIndex$ :
0 for minima, 1 for 1-saddles, $d-1$ for $(d-1)$-saddles, and $d$ for
 maxima.

Each topological feature of $\sublevelset{f}(\isovalue)$ 
(\emph{e.\,g.}, connected 
components, independent cycles, voids)
can be 
associated with a unique pair of critical points $(c, c')$, corresponding to 
 its 
\emph{birth} and \emph{death}.
Specifically, the Elder rule 
\cite{edelsbrunner09} states that critical points can be arranged according to 
this observation in a set of pairs, such that each critical point appears in 
only one pair $(c, c')$ such that $f(c) < f(c')$ and $\criticalIndex{c} = 
\criticalIndex{c'} - 1$. Intuitively, this rule implies that if two topological 
features of  $\sublevelset{f}(\isovalue)$ (\emph{e.\,g.}, two connected 
components) meet at a critical point $c'$, the \emph{youngest} feature 
(\emph{i.\,e.}, created last) \emph{dies}, favoring the \emph{oldest} one (\emph{i.\,e.}, created 
first). Critical point pairs can be visually represented by the 
\emph{persistence diagram}, noted $\persistenceDiagram{f}$, which embeds each 
pair to a single point in the 2D plane at coordinates $\big(f(c), f(c')\big)$, 
which respectively correspond to the birth and death of the associated 
topological feature (\autoref{fig:toyExample}). The \emph{persistence} of a 
pair, noted $\persistence(c, 
c')$, is then given by its height $f(c') - f(c)$.
It describes the lifetime in the range of the 
corresponding topological feature.

\paragraph{Wasserstein distance between persistence diagrams}

To cluster persistence diagrams, a first necessary 
ingredient is the notion of distance between them. 
Given two diagrams $\persistenceDiagram{f}$ and
$\persistenceDiagram{g}$, a pointwise distance 
can be introduced 
in the 2D birth/death space between   two points $a = (x_a, y_a) \in 
\persistenceDiagram{f}$ and $b = (x_b, y_b) \in \persistenceDiagram{g}$ by
\begin{equation}
\pointMetric{q}(a,b)=\left(|x_b-x_a|^2 + |y_b-y_a|^2\right)^{1/2} = \|a-b\|_2.
\label{eq_pointWise_metric}
\end{equation}

By convention, $\pointMetric{q}(a, b)$ is set to zero 
if both $a$ and $b$ exactly lie on the diagonal ($x_a = y_a$ and $x_b = y_b$).
The \emph{Wasserstein distance} between $\persistenceDiagram{f}$ and 
$\persistenceDiagram{g}$ can then be introduced as
\begin{equation}
    \wasserstein{q}\big(\persistenceDiagram{f}, \persistenceDiagram{g}\big) = 
\min_{\phi
\in \Phi} \left(\sum_{a \in \persistenceDiagram{f}} 
\pointMetric{q}\big(a,\phi(a)\big)^2\right)^{1/2},
\label{eq_wasserstein}
\end{equation}
where $\Phi$ is the set of all possible assignments $\phi$ mapping each 
point
$a \in \persistenceDiagram{f}$ to 
a point
$b 
\in \persistenceDiagram{g}$, 
or to 
its projection onto the diagonal, 
$(\frac{x_a+y_a}{2},\frac{x_a+y_a}{2})$, which denotes the removal of the 
corresponding feature from the assignment.
The Wasserstein distance can be computed by solving an 
optimal assignment problem, for which efficient approximation algorithms exist 
\cite{Bertsekas81, Kerber2016}.

It can often be useful to geometrically lift the 
Wasserstein metric by also taking into account
the geometrical layout of critical points \cite{soler2018}. Let 
$(c, c')$ be the 
critical point
pair corresponding to the point 
$a 
\in \diagram(f)$. Let  $p_{a}^\criticalPointMixing = \criticalPointMixing c' + 
(1 - 
\criticalPointMixing) c \in 
\range^d$ be their linear combination with coefficient $\criticalPointMixing 
\in 
[0, 1]$ in $\domain$. Our experiments (\autoref{sec:results})
only deal with
extrema, and we set $\criticalPointMixing$ to $0$ for minima and $1$ 
for maxima 
(to only consider the extremum's location).
Then, 
 the
geometrically lifted pointwise distance $\liftedMetric{2}(a, b)$ is 
given as
$\liftedMetric{2}(a, b) = 
  \sqrt{
  (1 - \geometricLifting) \pointMetric{2}(a, b)^2 
  + \geometricLifting ||p_a^\criticalPointMixing - 
p_b^\criticalPointMixing||_2^2}$.

\noindent
The parameter
$\geometricLifting \in [0, 1]$ 
quantifies the importance given to the geometry of  critical 
points 
and it must be tuned on a per application basis.

\paragraph{Fr\'echet mean of persistence diagrams}
Once a distance metric is established between topological signatures, 
a second ingredient is needed, namely the notion of barycenter, in order to 
leverage typical clustering algorithms.

Let $\diagramSpace$ be the space of persistence diagrams. 
The \needsVerification{discrete} \emph{Wasserstein barycenter} 
of a set $\{\persistenceDiagram{f_1}, \persistenceDiagram{f_2}, 
\dots, \persistenceDiagram{f_n}\}$ of persistence diagrams can  be 
introduced as the Fr\'echet mean of the set under 
the metric $\wasserstein{2}$.
It is 
the diagram $\diagramBarycenter$
that 
minimizes its 
distance to all the diagrams of the set (\emph{i.\,e.}, the minimizer of the so-called
Fr\'echet energy), that is,
$\diagramBarycenter = \argmin_{\diagram \in \diagramSpace} 
\sum_{i=1}^n \wasserstein{2}\big(\diagram, 
\persistenceDiagram{f_i}\big)^2$.
The computation of Wasserstein barycenters  involves a computationally 
demanding optimization problem, for which
the existence of at least one 
locally optimum solution has been shown by Turner et al. \cite{Turner2014}. 
Efficient algorithms have been proposed to solve this optimization problem 
\cite{lacombe2018}, including the progressive approach by Vidal et al. 
\cite{vidal_vis19}, which can return relevant approximations of Wasserstein 
barycenters, given some user defined time constraint $t_\mathrm{max}$, which is 
relevant for our urgent decision making context.

\paragraph{Topological clustering}
Once barycenters between topological signatures can be computed, traditional 
clustering algorithms, such as the $k$-means \cite{lloyd82},
can be revisited to support topological data representations.
Based on their 
efficient and progressive approach for Wasserstein barycenters, Vidal et al. 
\cite{vidal_vis19} revisit the $k$-means algorithm as follows.
The $k$-means is an iterative algorithm, where each \emph{Clustering} iteration 
is composed itself of two sub-routines: \emph{(i) Assignment} and \emph{(ii) 
Update}. Initially, $k$ cluster centroids $\diagramBarycenter_j$ ($j = 1,\ldots, 
k$) are initialized to $k$ diagrams $\diagram(f_i)$ from the input set.
Then, the \emph{Assignment} step consists of 
assigning each diagram $\diagram(f_i)$ to its closest centroid 
$\diagramBarycenter_{j(i)}$. This requires the computation of 
the Wasserstein distances $\wasserstein{2}$, of every diagram 
$\diagram(f_i)$ to all the 
centroids $\diagramBarycenter_j$.
Next, the \emph{Update} step consists of updating each 
centroid's location by placing it at the Wasserstein barycenter 
of its assigned diagrams  $\diagram(f_i)$. The algorithm 
continues these \emph{Clustering} iterations until convergence, that is,
until the assignments $i\mapsto j(i)$ between the diagrams and the $k$ 
centroids do not evolve anymore. Since Wasserstein barycenters can be approximated under user-defined 
time constraints with Vidal's approach \cite{vidal_vis19}, the above 
algorithm also supports time constraints (see \cite{vidal_vis19} for 
further details).
Of course, a larger time constraint will, in general, result in a better clustering of the input set of persistence diagrams.

\begin{figure*}
    \centering
    \includegraphics[width=.95\linewidth]{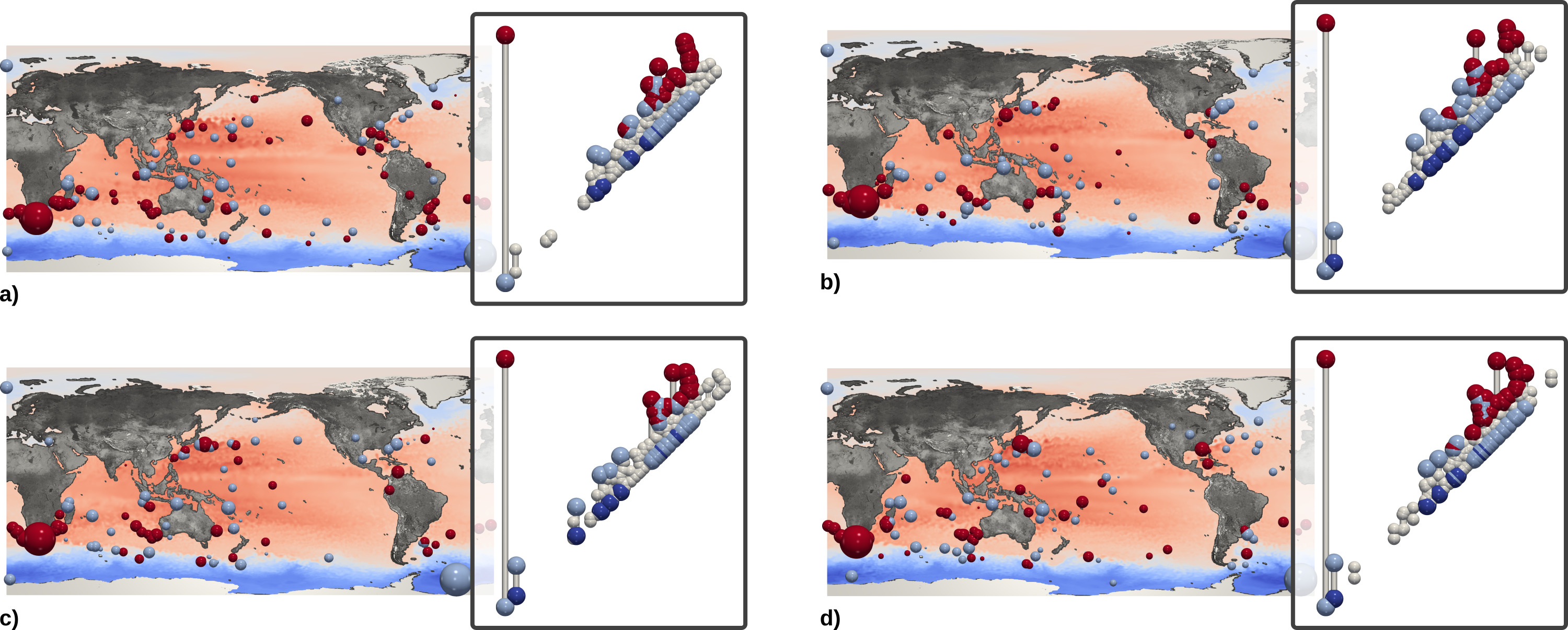}
    \caption{Clusters automatically identified by our topological clustering 
      approach ($t_\mathrm{max}$: 10 seconds) \jules{on the \emph{Sea Surface 
Height} data-set}. Left to right, top to bottom: pointwise mean 
    of each cluster.
    Inset diagram: cluster centroid computed by the 
        algorithm of Vidal et al. \cite{vidal_vis19} \jules{(in the 
diagrams, the X and Y axes denote the \emph{birth} and \emph{death} of the 
topological features, respectively)}. Barycenter extrema are 
      scaled in the domain by persistence \jules{ and colored by critical 
index} 
(spheres). \jules{In this example, the four clusters correspond to the four 
seasons.}}
    \label{fig:ssh}
\end{figure*}

\begin{figure*}
    \centering
    \includegraphics[width=.95\linewidth]{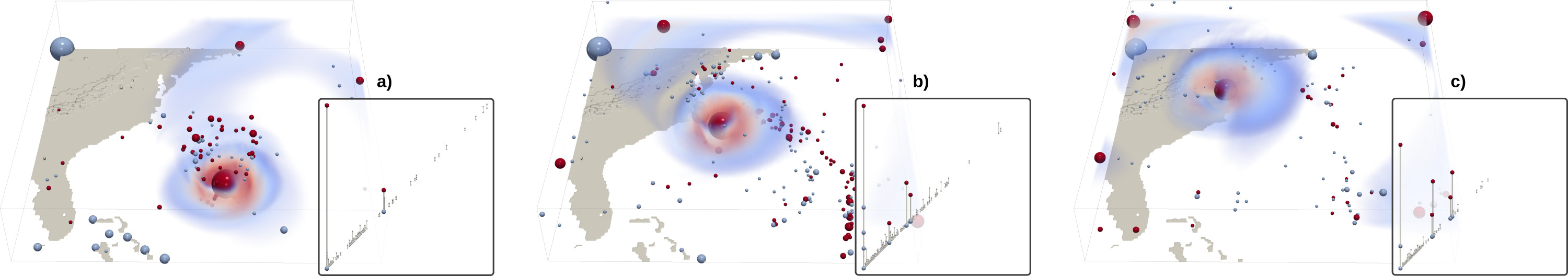}
    \caption{Clusters automatically identified by our topological clustering 
      approach ($t_\mathrm{max}$: 10 seconds) \jules{on the \emph{Isabel} 
data-set}. Left to right: pointwise mean of each 
        cluster. Inset diagram: cluster centroid computed by the 
        interruptible 
        algorithm of Vidal et al. \cite{vidal_vis19}  \jules{(in the 
diagrams, the X and Y axes denote the \emph{birth} and \emph{death} of the 
topological features, respectively)}. Barycenter extrema are 
      scaled in the domain by persistence \jules{ and colored by critical 
index} 
(spheres). \jules{In this example, the three clusters correspond to the three 
hurricane configurations (from left to right: formation, drift and landfall).}}
    \label{fig:isabel}
\end{figure*}

\subsection{Statistical scores}
The previously described method assumes that the number of clusters $k$ is known \emph{a priori}.
If the number of clusters is not known in advance, so-called \emph{information criteria} can be used to select a number of clusters \emph{a posteriori} after the $k$-means algorithm has been applied for several values of $k$.

In our application, we will use the Akaike Information criterion (AIC, \cite{Akaike1973a,Akaike1973b}) and the Bayesian information criterion (BIC, \cite{Schwarz1978}), which are based on the minimization of a score function of the form
\begin{align}
    \mathrm{IC}(k) = 2\,L(k) + p(k),\label{eq:IC}
\end{align}
where $L(k)$ is the value of the log-likelihood function of the clustering result, when $k$ clusters are detremined, and the term $p(k)$ penalizes the number of parameters differently for AIC and BIC.
The criteria can be interpreted as a way to balance the goodness of fit (represented by the log-likelihood function) and the number of parameters:
on the one hand, the goodness of fit is minimal if the number of clusters and the number of data points coincide, but the number of parameters is high in this situation.
On the other hand, if the number of clusters is minimal, then the goodness of fit is, generally, very large.
The minimum value of the information criterion will, consequently, be somewhere inbetween.

Under the so-called \emph{identical spherical assumption} (see \cite{Pelleg2000}), it can be shown (originally, for data from a Euclidean space) that the log-likelihood term has the form
\begin{align}
L &= \sum_{j=1}^k n_j\,\log{n_j} - n\,\log{n} - \frac{n\,d}{2}\,\log(2\,\pi\,\hat{\sigma}^2) - \frac{d}{2} \,(n-k),\label{eq:loglik}
\end{align}
where $n_j$ is the number of diagrams mapped to the centroid $\diagramBarycenter_j$, $n$ is the total number of diagrams, $d$ is the dimension of $\diagramSpace$, and $\hat{\sigma}$ is an estimation of the in-cluster variance, for example,
\begin{align}
\hat{\sigma}^2 &= \frac{1}{d\,(n-k)} \sum_{i=1}^n \wasserstein{}(\diagram(f_i), \diagramBarycenter_{j(i)})^2.\label{eq:variance}
\end{align}
Since the dimension of $\diagramSpace$ cannot be easily determined, we choose a value for $d$ in our prototype implementation such that the information criteria show the expected behavior (approximately convex, being monotonically decreasing for small $k$ and monotonically increasing for large $k$).

The penalty term $p$ in \eqref{eq:IC} for the AIC is given by
\begin{align}
p_\mathrm{AIC}(k) = 2\,k\,d,
\end{align}
whereas for the BIC it is given by
\begin{align}
p_\mathrm{BIC}(k,N) = k\,d\,\log(N)
\end{align}
(cf.\ \cite{EfronHastie2016}, Sect.~13.3),
where the term $k\,d$ encodes the number of effective parameters of the statistical model, particularly, the $d$ coordinates of the $k$ cluster centers.
Note that for a comparison of different clusterings of a fixed data set, $p_\mathrm{AIC}$ and $p_\mathrm{BIC}$ do indeed only depend on $k$, since both the dimension $d$ of the underlying space as well as the number $N$ of samples is constant.

\section{Prototype Implementation}
This section details the implementation of our 
prototype.

\subsection{Topological clustering}
For each ensemble data set, given a user time constraint $t_\mathrm{max}$, we 
systematically run the progressive topological clustering algorithm of Vidal et 
al.\ \cite{vidal_vis19} for a range of $k$ values (typically, $1$ to $10$). For 
this, we used the companion C++ implementation provided by Vidal et al.\
\cite{vidal_vis19}\jules{, available in the Topology ToolKit (TTK) \cite{ttk17}}. Since the computation is independent for distinct $k$ 
values, this step can be trivially parallelized (one
$k$-clustering per process/thread).

\subsection{Statistical scores}
Once the clustering has been performed for different values of $k$, the computation of the statistical scores (AIC and BIC) is straight-forward if the Wasserstein distances of each persistence diagram to its nearest centroid are extracted from the clustering process.
Inserting these distances in \eqref{eq:variance} results in an estimation of the in-cluster variance, which can then be used in \eqref{eq:loglik} to compute the value of the log-likelihood function.
Combined with the computation of the penalty terms $p_\mathrm{AIC}$ and $p_\mathrm{BIC}$, one obtains a value of the statistical score for the given clustering.

\begin{figure*}
    \centering\footnotesize
    \includegraphics[width=0.25\linewidth]{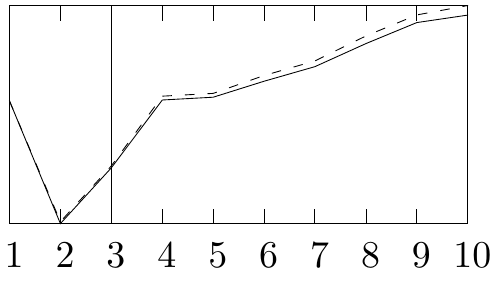}    
    \hspace{1cm}
    \includegraphics[width=0.25\linewidth]{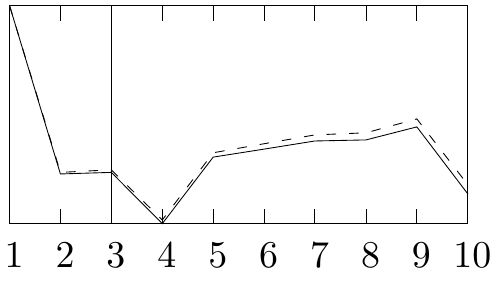}    
    \hspace{1cm}
    \includegraphics[width=0.25\linewidth]{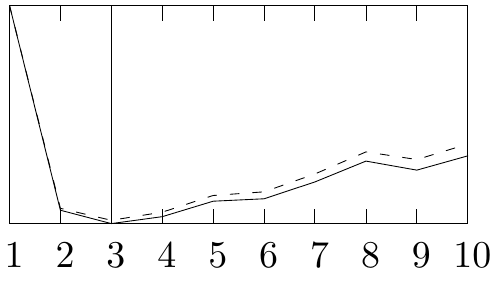}\\
    (a) \emph{Gaussians} ensemble\\
    \includegraphics[width=0.25\linewidth]{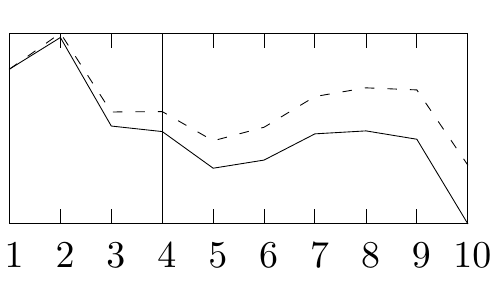}    
    \hspace{1cm}
    \includegraphics[width=0.25\linewidth]{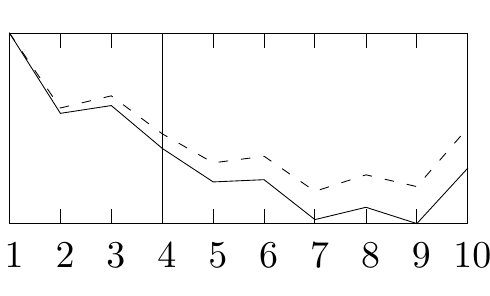}    
    \hspace{1cm}
    \includegraphics[width=0.25\linewidth]{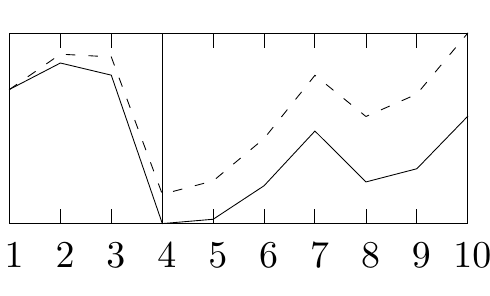}\\
    (b) \emph{Sea Surface Height} ensemble\\
    \includegraphics[width=0.25\linewidth]{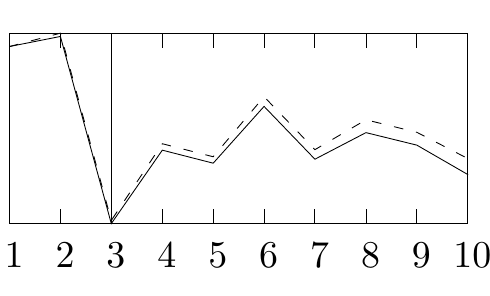}    
    \hspace{1cm}
    \includegraphics[width=0.25\linewidth]{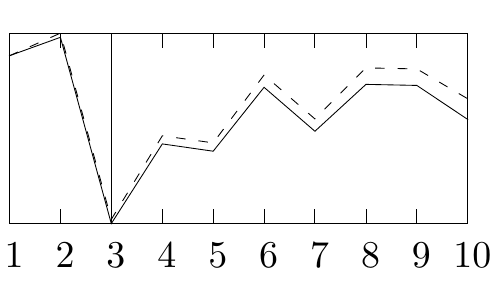}    
    \hspace{1cm}
    \includegraphics[width=0.25\linewidth]{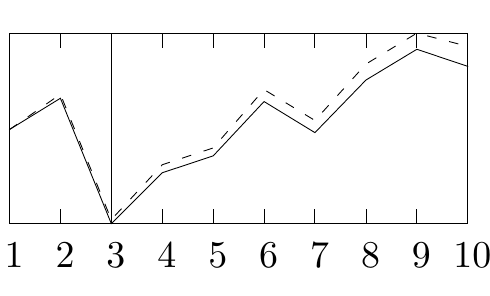}\\
    (c) \emph{Isabel} ensemble
    \caption{Values of the AIC (solid line) and BIC (dashed line) for $k=1, 
\ldots, 10$ for the three ensemble data sets for $t_\mathrm{max}=1\,\mathrm{s}, 
10\,\mathrm{s}, 100\,\mathrm{s}$ (left-to-right).
\jules{The values have been normalized to the value for $k=1$ for each diagram. Therefore 
the Y axes are not labeled. The X axes denote the number of clusters.}}
    \label{fig:scores}
\end{figure*}

\section{Results}
\label{sec:results}
This section presents experimental results obtained 
with a C++ implementation.
The  input persistence diagrams were computed with the algorithm 
by Gueunet et al. 
\cite{gueunet_tpds19}, 
which is available in the Topology 
ToolKit (TTK) 
\cite{ttk17}.

Our experiments were performed on a variety of simulated and acquired 2D and 3D 
ensembles, taken from Favelier et al. \cite{favelier2018}, following the 
experimental setup of Vidal et al. \cite{vidal_vis19}.
The \emph{Gaussians} ensemble contains 100 2D synthetic 
noisy members, with 
3 patterns of Gaussians.
The \emph{Sea Surface Height} ensemble 
(\autoref{fig:ssh}) is composed of 
48 observations taken in January, April, July and October 2012 
(\href{https://ecco.jpl.nasa.gov/products/all/}{https://ecco.jpl.nasa.gov/products/all/}). 
Here, the 
features of 
interest are the center of  eddies, which can be 
reliably estimated with 
height extrema. Thus, both the diagrams involving the minima and maxima are 
considered and 
independently processed by our algorithms.
Finally, the \emph{Isabel} data set (\autoref{fig:isabel}) is 
a volume ensemble of 12 members, showing key time 
steps (formation, 
drift and landfall) in the simulation of the Isabel hurricane 
\cite{scivisIsabel}. In this example, the eyewall of the hurricane is typically 
characterized by high wind velocities, well captured by velocity maxima. Thus 
we only consider diagrams involving maxima.
Unless stated otherwise, all 
results were obtained by considering the Wasserstein metric $\wasserstein{2}$ 
based on the original pointwise metric in \eqref{eq_pointWise_metric} without 
geometrical lifting (\emph{i.\,e.}, $\geometricLifting = 0$, 
\autoref{sec_background_tda}).

In \autoref{fig:scores}, we have depicted the values of the statistical score functions for these three data sets for three different values of $t_\mathrm{max}$, where the number of clusters is characterized as the minimizer of the score functions.
We have marked the number of clusters, determined with the most accurate clustering (that is, $t_\mathrm{max}=100\,\mathrm{s}$) with a vertical line.
We observe that for the less accurate clusterings ($t_\mathrm{max}=1\,\mathrm{s}, 10\,\mathrm{s}$), we may obtain either a just slightly different number of clusters (Gaussians ensemble) or a number nearly doubling the optimal number of clusters (Sea Surface Height ensemble).
This might seem to be a drawback with regard to the application of urgent 
decision making, where small values of $t_\mathrm{max}$ are desirable.
\jules{However, in practice, when comparing the identified clusters with the 
crisis situation in reality to determine the most likely outcome, it is very 
helpful if the number of clusters is much lower than the number of ensemble 
simulations.
This is still the case both for slightly different numbers of clusters and also for a twice as high number of clusters.
Additionally, when determining the number of clusters in time-varying ensemble 
simulations, as described in the introduction, it is especially interesting if 
the number of clusters changes at a specific time step.
We expect that these changes will also take place with the less accurate numbers
of clusters.
Of course, this will be analyzed in more detail in the future, when the 
presented method will be applied to data sets from the pilot applications used 
in the VESTEC project \cite{vestec}.}

\autoref{fig:ssh} shows our results for the \emph{Sea Surface 
Height} ensemble, where our statistical analysis estimates an optimal number of 
clusters of $k=4$ and where the topological clustering \cite{vidal_vis19}
automatically identifies 
four clusters, corresponding to the four seasons: 
winter, spring, summer, fall (left-to-right, top-to-bottom). 
As shown in the insets, each season leads to a visually distinct centroid 
diagram.

As discussed by Vidal et al. \cite{vidal_vis19}, geometrical lifting is 
particularly important in applications where feature location bears a meaning, 
such as the Isabel ensemble 
(\autoref{fig:isabel}). For this example, 
our statistical analysis estimates an optimal number of clusters of $k=3$ and
the clustering algorithm with 
geometrical lifting \cite{vidal_vis19} automatically identifies the right 
clusters, corresponding to the three states of the hurricane (formation, drift 
and landfall).

\section{Conclusion}
Motivated by urgent decision making applications, which require the clustering of ensemble simulation outputs for the determination of different crisis scenarios, we proposed a statistical technique to determine the number of clusters based on a recently published progressive clustering method for so-called persistence diagrams.
We presented a proof-of-concept prototype implementation, which provided meaningful results for real-world ensemble data sets.
In our upcoming research, we will incorporate the parameter selection within 
the clustering approach directly based on this prototype.
\jules{Using Paraview Catalyst \cite{catalyst_ayachit_2015, catalyst_tuto_ttk}, 
our approach can easily be integrated into any simulation
code. It then allows to carry out \emph{in-situ} clustering operations on a 
statistically determined number of clusters at chosen iterations of the 
simulation, while respecting the time constraints of an urgent decision making 
situation.}
Furthermore, in the context of the European project VESTEC \cite{vestec}, we 
will apply our 
approach to other real-life use cases (wildfire, mosquito-borne diseases, space 
weather) and, especially, in an in-situ context to allow for an interaction of 
the decision maker with the ensemble simulations.

\section*{Acknowledgments}
The authors would like to thank 
the anonymous reviewers for their thoughtful remarks and suggestions.

\bibliographystyle{abbrv-doi}
\bibliography{paper}

\end{document}